\documentclass[aps,prx,twocolumn,nofootinbib,superscriptaddress,longbibliography]{revtex4-2}

\usepackage{amsmath,amsfonts,amssymb,bm}
\usepackage{graphicx}
\usepackage{xcolor}
\usepackage[colorlinks=true, linkcolor=blue, citecolor=blue, urlcolor=blue, breaklinks=true]{hyperref}
\usepackage{physics}
\usepackage{dsfont}
\usepackage{bbold}
\usepackage{soul}

\newcommand{\figpanel}[1]{\textbf{\textsf{#1}}}

\usepackage{etoolbox}
\patchcmd{\section}
  {\centering}
  {\raggedright}
  {}
  {}
\patchcmd{\subsection}
  {\centering}
  {\raggedright}
  {}
  {}

\begin{document}

\title{Finite-temperature criticality through quantum annealing}

\author{Gianluca Teza}
\email{teza@pks.mpg.de}
\affiliation{Max Planck Institute for the Physics of Complex Systems, N\"othnitzer Str.~38, 01187 Dresden, Germany}

\author{Francesco Campaioli}
\email{francesco.campaioli@rmit.edu.au}
\affiliation{Department of Physics, School of Science, RMIT University, Melbourne, Victoria, Australia}
\affiliation{Dipartimento di Fisica e Astronomia G. Galilei, Universit\'a degli Studi di Padova, 35131 Padova, Italy}

\author{Marco Avesani}
\email{marco.avesani@unipd.it}
\affiliation{Dipartimento di Ingegneria dell’Informazione, Universit\'a degli Studi di Padova, 35131 Padova, Italy}
\affiliation{Padua Quantum Technologies Research Center, Universit\`a degli Studi di Padova, via Gradenigo 6B, IT-35131 Padova, Italy}

\author{Oren Raz}
\email{oren.raz@weizmann.ac.il}
\affiliation{Department of physics of complex systems, Weizmann institute of science, Rehovot, Israel, 76100}

\date{\today}

\begin{abstract}
Critical phenomena at finite temperature underpin a broad range of physical systems, yet their study remains challenging due to computational bottlenecks near phase transitions. Quantum annealers have attracted significant interest as a potential tool for accessing finite temperature criticality beyond classical reach, but their utility in precisely resolving criticality has remained limited by noise, hardware constraints, and thermal fluctuations. Here we overcome these challenges, showing that careful calibration and embedding allow quantum annealers to capture the full finite-temperature critical behavior of the paradigmatic two-dimensional Ising ferromagnet. By tuning the energy scale of the system and mitigating device asymmetries, we sample effective Boltzmann distributions and extract both the critical temperature and the associated critical exponents. Our approach opens the study of equilibrium and non-equilibrium critical phenomena in a broad class of systems at finite temperature.
\end{abstract}

\maketitle

Finite-temperature critical phenomena refer to the dramatic and universal changes that occur in a physical system as it approaches a phase transition at a non-zero temperature~\cite{goldenfeld2018lectures}. 
These phenomena arise across a wide range of settings, from common fluids to extreme conditions. Examples include the fluid critical point where liquid and gas become indistinguishable, the Curie temperature marking the loss of ferromagnetism~\cite{chaikin1995principles}, and the $\lambda$-point below which helium exhibits superfluidity~\cite{tilley1990superfluidity}.
At criticality, properties like correlation length, susceptibility, and heat capacity diverge or become anomalously large. These divergences follow universal scaling laws, defined by critical exponents that depend only on broad features like dimensionality and symmetry, rather than microscopic details~\cite{cardy1996scaling}. 
Understanding these phenomena is essential not only for theory but also for applications, where finite-temperature criticality underpins engineered responses such as magnetoresistance~\cite{tokura2000colossal}, superconductivity~\cite{fradkin2015colloquium}, and conductivity modulation across Mott transitions~\cite{imada1998metal}, key for advanced electronic devices.

Decades of progress in theoretical and computational physics have yielded powerful methods for studying critical phenomena, from the celebrated renormalization group theory~\cite{wilson1974renormalization} to advanced Monte Carlo algorithms~\cite{swendsen1987nonuniversal}. Yet, these methods face persistent challenges. Critical slowing down limits sampling efficiency near critical points~\cite{Langfeld2022}, the sign problem hinders simulations of frustrated and fermionic systems~\cite{li2024semigroup}, and tensor network approaches struggle with entanglement growth in higher dimensions~\cite{PhysRevB.109.235102}. These limitations have fueled the exploration of alternative strategies for studying finite-temperature criticality in complex systems.

Recently, quantum computing and simulation platforms have emerged as promising alternatives, with the expectation that quantum coherence and entanglement could open to regimes where classical methods struggle~\cite{altman2021quantum,georgescu2014quantum}. Digital quantum processors have been used to study quantum phase transitions and non-equilibrium critical dynamics~\cite{keesling2019quantum}, while analog platforms based on cold atoms or trapped ions have successfully realized spin models exhibiting critical behavior~\cite{zhang2017observation}. 

A particular platform that is gaining attention are quantum annealers. These are analog quantum devices designed to find low-energy configurations of classical spin systems by adiabatically evolving toward a problem-specific Ising Hamiltonian~\cite{johnson2011quantum}. Operating at finite temperature, they can approximately sample from Boltzmann distributions~\cite{amin2008effect, benedetti2016estimation, shibukawa2024boltzmann}, a feature that has inspired efforts to repurpose them as simulators for the thermodynamics of complex classical and quantum systems~\cite{buffoni2020thermodynamics}. Recent studies have shown that quantum annealers can capture features of classical and quantum phase transitions~\cite{king2018observation, heim2015quantum,hearth2022quantum,sathe2025classical}, highlighting their potential for exploring equilibrium statistical mechanics and finite-temperature criticality.
\begin{figure*}[t]
    \centering
    \includegraphics[width=\linewidth]{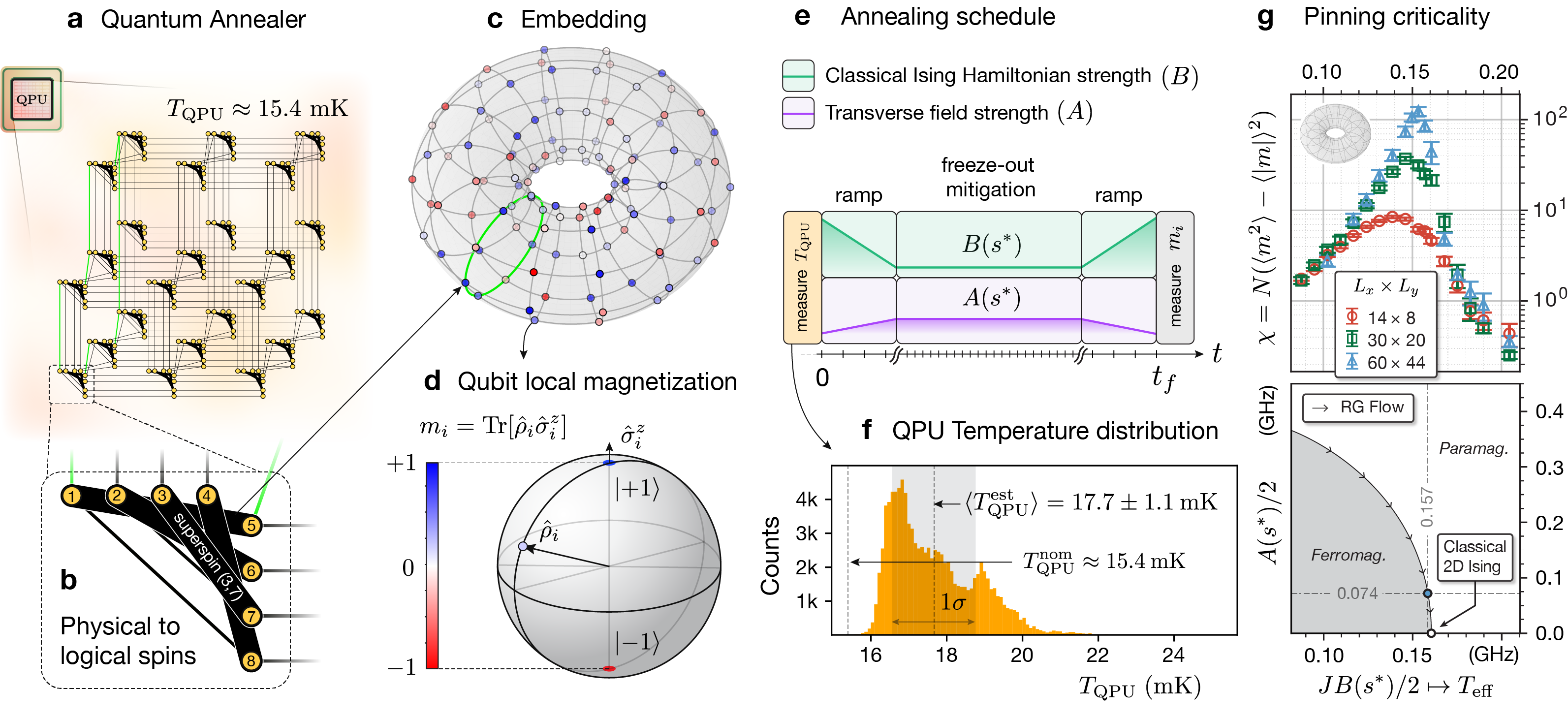}
    \caption{\textbf{Overview of the embedding and thermal sampling calibration.} \textbf{\textsf{a}}, Quantum processing unit (QPU) of D-Wave's Advantage system, consisting of a network of qubits cooled to a declared temperature $T_\mathrm{QPU}\approx 15.4\:\mathrm{mK}$, here shown for the embedding of the 2D Ising ferromagnet with toroidal topology. \textbf{\textsf{b}}, QPU's qubit pairs are strongly coupled to form logical qubits (\textit{superspin}), enabling periodic boundary conditions. 
    \textbf{\textsf{c}}, Toroidal $8 \times 14$ 2D Ising ferromagnet associated with the embedding of panel~\figpanel{a},~\figpanel{b}. A geodesic on the torus is highlighted in green and shown on the QPU graph. 
    \textbf{\textsf{d}}, Local qubit state $\hat{\rho}_i$ of site $i$ on the torus, with local magnetization $m_i$, represented on the Bloch sphere. \textbf{\textsf{e}}, Hardware-encoded annealing schedule: The functions $A(s)$ and $B(s)$ determine the transverse and longitudinal energy scales, respectively, as a function of the annealing parameter $s$. Sampling at a fixed intermediate point $s^*$ yields equilibrium distributions consistent with a Boltzmann ensemble at some effective temperature $T_\mathrm{eff}$ while preventing the system from getting stuck in an out-of-equilibrium state (\textit{freeze-out}). \textbf{\textsf{f}}, Histogram of single-qubit temperature calibration measurements necessary to capture deviations of $T_\mathrm{eff}$. \textbf{\textsf{g}}, Finite temperature criticality pinned by the magnetic susceptibility $\chi$, via the mapping between annealing parameters and the effective temperature $T_\mathrm{eff}$, here represented above a cartoon of the ferromagnet phase diagram.
    }
    \label{fig:embedding}
\end{figure*}

However, despite their promise, accurately pinning finite-temperature criticality with quantum annealers remains difficult, due to several challenges. Near the critical point, the system's response time to external changes diverges, demanding impractically long annealing durations~\cite{KingCoherent2022}. Boltzmann sampling is also affected by fluctuations in the temperature of the quantum processing unit (QPU), illustrated in Fig.~\ref{fig:embedding}~\figpanel{a}, which is influenced by both external factors and the specific choice of annealing schedules~\cite{benedetti2016estimation}. Furthermore, hardware constraints such as connectivity, coupling noise, and schedule-induced biases affect the reproducibility and reliability of measured observables~\cite{mishra2022quantum}.
Indeed, a recent study has highlighted the challenges of using quantum annealers to probe universal criticality and finite-size scaling when mapping the phase diagram of classical Ising systems of up to 144 spins~\cite{sathe2025classical}.

In this work, we tackle these challenges and definitively demonstrate that quantum annealers can be carefully programmed to precisely resolve finite-temperature criticality in complex many-body systems. To benchmark the quality of the approach, we leverage the two-dimensional (2D) ferromagnetic Ising model, thanks to its exactly known critical point~\cite{onsager1944crystal}. Magnetic leakage, hardware asymmetries, and edge effects are mitigated through a tailored embedding, i.e., mapping between the simulated system and the physical architecture of the annealer, depicted in Fig.~\ref{fig:embedding}~\figpanel{b},~\figpanel{c}. A range of effective temperatures is simulated by varying the energy scale of the model, calibrating the QPU's temperature before each anneal (as shown in Fig~\ref{fig:embedding}~\figpanel{e},~\figpanel{f}).
Altogether, this enables full use of the QPU network to embed lattices with more than 2500 spins.
From local observables (Fig.~\ref{fig:embedding}~\figpanel{d}) and two-point correlations---magnetization and susceptibility (Fig.~\ref{fig:embedding}~\figpanel{g})---we extract the critical temperature and the full set of critical exponents. 

The ability to  measure and control all the relevant parameters of an exactly solvable many body model is extremely useful. We demonstrate some of the corresponding implications in two separated measurements: in the first, we track the magnetization during a non-equilibrium relaxation of the system after a sudden quench through the phase transition in the presence of a small longitudinal and transverse magnetic fields. This demonstrates the acceleration gained by the {presence of a} transverse field. In a second demonstration, we initiate the system at the metastable state associated with all spins pointing against the longitudinal magnetic field. We then measure the probability of the system to tunnel through the barrier between the metastable and stable states. Both the barrier height and the transverse field can be controlled independently, shedding light on a key property of quantum annealers, i.e., their ability to tunnel through energy barriers to reach lower-energy states.      

Our approach establishes a systematic and scalable method for studying finite-temperature criticality using quantum annealers, applicable to a broad class of classical and quantum systems with embedded Ising representations. By leveraging the tunability of couplings and the relatively high connectivity available on current annealing architectures---now beyond~15 couplings per spin---our methodology extends to frustrated magnets~\cite{zhao2024quantum}, spin glasses~\cite{KingCritical2023}, and constrained models relevant to optimization~\cite{bian2016mapping}, and lattice gauge theories~\cite{rahman2021su2}. Moreover, the same programmable framework can be adapted to investigate non-equilibrium dynamics and quenched disorder, offering a path for the exploration of critical phenomena in regimes that are challenging for classical methods.

\section*{Embedding the 2D Ising Model}
\noindent
In the 2D Ising model, classical spins on a square lattice are coupled with a nearest-neighbors interactions. The model has a disordered phase above its critical temperature. In this phase, the probability to find a spin in each of its state is equal. Below the critical temperature the system has a spontaneously broken symmetry, and most spins points in the same direction. The 2D Ising model is a natural candidate for our purpose due to its well-understood equilibrium properties and its prototypical finite-temperature phase transition. To use this model as a benchmark for a quantum annealer, we begin by embedding the classical 2D Ising ferromagnet onto the hardware graph of the D-Wave Advantage system, which is based on the Pegasus topology~\cite{dwave_docs}.  
\begin{figure}[t]
    \centering
    \includegraphics[width=0.46\textwidth]{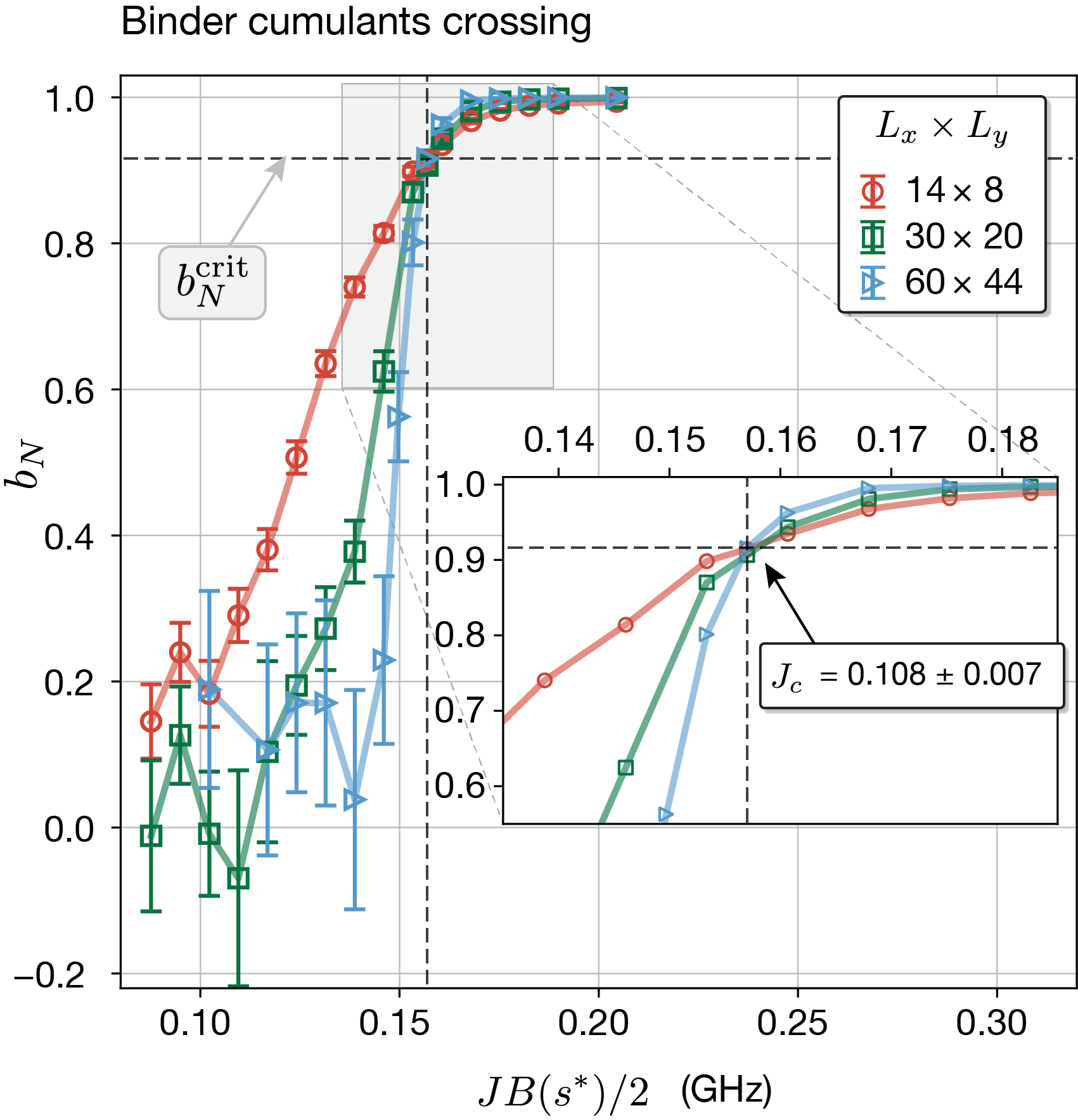}
    \caption{\textbf{Pinning finite-temperature criticality.}
    Binder cumulant $b_N(J)$ for three system sizes ($N = 112$, $600$, $2640$) as a function of the effective ferromagnetic coupling $J$. 
    The curves cross at $J_c = 0.108\pm0.007$, indicating the presence of a critical point in the thermodynamic limit (inset). 
    The crossing value $b_N(J_c) = 0.91\pm0.04$ matches the known value for a 2D Ising model with periodic boundary conditions and aspect ratio $L_x/L_y \approx 3/2$.
    This validates the effectiveness of the sampling protocol in recovering critical behavior.
    }
    \label{fig:binder_cumulant}
\end{figure}

Directly realizing a regular square lattice with periodic boundary conditions (PBCs) on the Pegasus graph requires careful mapping, as the native qubit connectivity is both sparse and non-planar. We employ a tiling-based embedding strategy that groups strongly coupled physical qubits into logical ``superspins'', as shown in Fig. \ref{fig:embedding}~\figpanel{a}, \figpanel{b}. This approach---used in several previous studies to emulate lattice connectivity and enhance coherence---enables each superspin to behave as a single quantum spin variable, with its internal ferromagnetic couplings set to be significantly stronger than the interactions between superspins~\cite{bunyk2014architectural,king2022scaling}. It also allows us to impose effective toroidal boundary conditions, shown in Fig.~\ref{fig:embedding}~\figpanel{c}, and mitigate edge effects that would otherwise distort the scaling analysis in finite systems~\cite{kaneda2001finite}.

We implement ferromagnetic couplings of high amplitude (15--20 times stronger than those used as the interaction between different superspins) to enforce coherence within each superspin. However, such strong couplings can induce magnetic leakage, leading to systematic biases in effective fields experienced by neighboring spins. To correct for these biases, we follow a calibration protocol that introduces local field offsets determined via single-qubit benchmarking procedures~\cite{chern2023tutorial}. This compensation ensures consistency across different experimental runs and mitigates embedding-related noise.
\begin{figure*}[t]
    \centering
    \includegraphics[width=\linewidth]{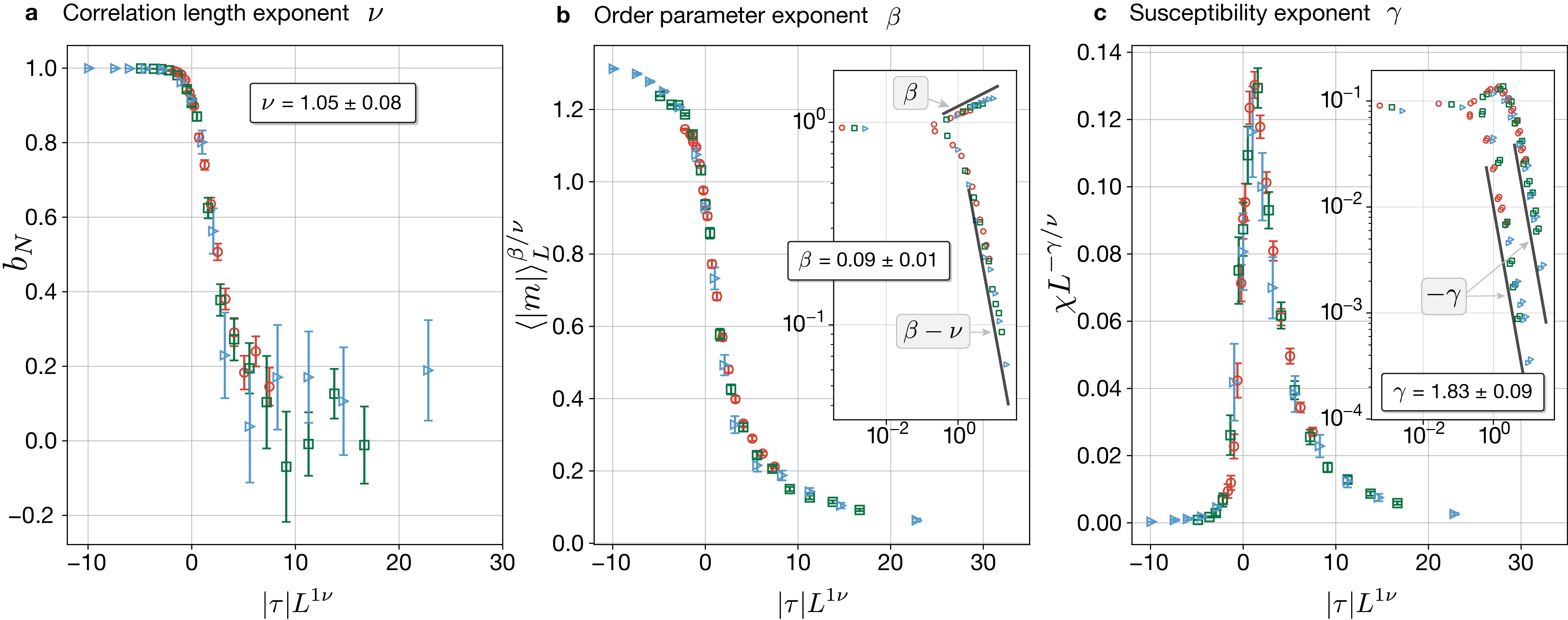}
    \caption{\textbf{Recovering the critical exponents.}
    \textbf{\textsf{a}}, Plotting the Binder cumulant against the reduced temperature $\tau=J_c/(J-J_c)$ rescaled by $L^{\nu}$, where $\nu$ is the correlation exponent and $L=\sqrt{L_x\times L_y}$.
    The plot shows an excellent collapse around the critical region $t=0$ for the extracted value $\nu=1.05\pm0.08$ , in agreement with the expected value for a 2D Ising ferromagnet at criticality.
    \textbf{\textsf{b}}, Collapse of the rescaled order parameter $\left<|m|\right>L^{\beta/\nu}$ shows an excellent agreement of the data for the three different system sizes $N$, yielding an extracted order parameter exponent $\beta=0.09\pm0.01$.
    Visualization in log-scale shows perfect agreement with the power-law behavior $\sim\beta-\nu$ ($\sim \beta)$ for the negative (positive) branch. 
    \textbf{\textsf{c}}, Collapse of the rescaled susceptibility $\chi L^{-\gamma/\nu}$ shows a strong agreement with the different system sizes and a great compatibility of the extracted exponent $\gamma=1.83\pm0.09$ with the expected value. Visualization in log-scale shows perfect agreement with the power-law behavior $\sim-\gamma$ for both $t\lessgtr0$ branches.
    }
    \label{fig:exponents}
\end{figure*}

Using this method, we successfully embed  rectangular lattices with system sizes up to 2640 spins (e.g., $60 \times 44$), while maintaining the aspect ratio $L_x/L_y \approx 3/2$ for consistency in scaling analysis. Our largest systems span nearly the full available qubit count on the Advantage system. By constructing multiple such embeddings with randomized gauge transformations (spin-flip symmetries \cite{mattis1976solvable}), we average out spatial asymmetries in hardware behavior~\cite{boothby2020next}.

This embedding strategy forms the foundation for reliable thermal sampling, enabling us to later interpret critical behavior through finite-size scaling and cumulant analysis of sampled spin configurations.

\section*{Finite-Temperature Sampling}
\noindent
To study equilibrium properties at finite temperature using a quantum annealer, we exploit the device’s ability to sample from thermally populated states near the end of its annealing schedule. The D-Wave system implements a transverse-field Ising model with a time-dependent Hamiltonian:
\begin{equation}
\hat{H}(s) = -\frac{A(s)}{2} \sum_i \hat{\sigma}^x_i + \frac{B(s)}{2} \left[ \sum_i h_i \hat{\sigma}^z_i + \sum_{i,j} J_{ij} \hat{\sigma}^z_i \hat{\sigma}^z_j \right],
\end{equation}
where $\hat{\sigma}^x_i$ and $\hat{\sigma}^z_i$ are Pauli matrices acting on qubit $i$, $h_i$ and $J_{ij}$ are programmable local fields and couplings, and $s \in [0,1]$ parametrizes the annealing schedule. The functions $A(s)$ and $B(s)$ control the transverse and longitudinal energy scales and are fixed by the hardware.

At the end of the anneal, the transverse field ideally vanishes and the system is governed by the classical Hamiltonian:
\begin{equation}
H(\{\sigma_i\}) = \frac{B(1)}{2} \left[ \sum_i h_i \sigma_i + \sum_{i,j} J_{ij} \sigma_i \sigma_j \right],
\end{equation}
where $\sigma_i = \pm 1$ and $B(1)$ is the final value of the longitudinal energy scale with anneal parameter $s=1$. If the system remains in contact with thermal environment---the fridge, QPU  (Quantum Processing Unit) itself, and other sources---for long enough time, then the sampled spin configurations at the end follow a Boltzmann distribution \cite{NelsonThermalGibbs2022}
\begin{equation}
P(\{\sigma_i\}) \propto \exp\left( -\frac{H(\{\sigma_i\})}{k_B T_q} \right),
\end{equation}
where $T_q$ is the physical temperature of the QPU.

In practice, thermal sampling near \( s = 1 \) is complicated by the so-called freeze-out phenomenon \cite{king2020scaling,MarshallPausing2019}, where the dynamics slow down dramatically and the system becomes trapped in metastable configurations. While the precise onset of freeze-out is not directly encoded in hardware parameters, the annealing schedule shown in Fig.~\ref{fig:embedding}~\figpanel{e}---which defines the evolution of the transverse and longitudinal energy scales via $A(s)$ and $B(s)$---constrains the thermal accessibility of classical configurations. To mitigate freeze-out, we halt the anneal at a control parameter value  $s^*=0.5$, such that $A(s^*)/B(s^*)\approx0.05$~\cite{dwave_docs}, ensuring the system remains in a dynamic regime while still allowing for effectively classical readout. Sampling at this intermediate 
point yields distributions consistent with thermal equilibrium~\cite{lokhov2018optimal}.
The choice of \( s^* \) is validated a posteriori via Binder cumulant crossings and critical scaling, as described below.

A second important issue is temperature calibration \cite{BenedettiTemperature2016,raymond2016global}. The nominal operating temperature of the QPU ($T^{\textrm{nom}}_\textrm{QPU} \approx 15.4$ mK) can fluctuate significantly between experimental runs. These fluctuations are caused by both environmental noise and internal operational procedures, including device programming, readout cycles, and idle state drift.
As shown in Fig.~\ref{fig:embedding}~\figpanel{f}, these variations are directly measurable: The histogram displays the distribution of effective QPU temperatures inferred from repeated calibration steps. To compensate for these fluctuations, we perform an in situ temperature estimation before each batch of samples.
Specifically, we program single-qubit thermalization problems and extract a maximum-likelihood estimate of their freeze-out temperatures $T^{\textrm{est}}_{\textrm{FO}}$ based on excitation probabilities~\cite{raymond2016global}.
The value of the anneal control parameter at which the dynamics of this problem freezes out is known ($s_{\textrm{FO}}=0.612$), and allows to assess a declared effective temperature at freeze-out of $T^{\textrm{nom}}_{\textrm{FO}}=2 k_b T^{\textrm{nom}}_{\textrm{QPU}}/B(s_{\textrm{FO}})\approx 0.164$ ($k_B=1$ units).
The inferred correction factor $\delta = T_\textrm{FO}^\text{est} / T_\textrm{FO}^\text{nom}$ is then used to rescale the classical energy scale $J$ across experiments, preserving consistency in the effective sampling temperature.
Altogether, this allows us to estimate an average effective temperature of the QPU throughout all the performed experiments as $\left< T_\textrm{QPU}^{\textrm{est}} \right>=17.7 \pm 1.1\ \textrm{mK}$, where the associated error is the variability of the temperature across the experiments.

To our knowledge, this form of real-time temperature compensation has not been applied in previous quantum annealing studies of Boltzmann sampling. Combined with early-termination protocols, it enables consistent finite-temperature sampling across multiple runs, device conditions, and---most importantly---system sizes. Indeed, these steps are fundamental to fully exploit the 5000+ qubits of the QPU \cite{dwave_docs}, allowing us to embed sizes considerably larger than those used so far for finite-temperature analyses \cite{ParkFrustrated2022,hearth2022quantum,sathe2025classical}.

\section*{Pinning Criticality via Binder Cumulants}
\noindent
To determine whether our annealing protocol successfully prepares the system at the critical point of the 2D Ising model, we analyze the Binder cumulant $b_N(J)$ as a function of the ferromagnetic coupling strength $J$ \cite{binder1981finite}. The Binder cumulant is defined via moments of the total magnetization $M = \sum_i \sigma_i$:
\begin{equation}
b_N(J) = \frac{1}{2} \left( 3 - \frac{\langle M^4 \rangle}{\langle M^2 \rangle^2} \right),
\end{equation}
where the averages are taken over samples generated for a system of $N$ spins at fixed $J$. This quantity is sensitive to the shape of the magnetization distribution: it approaches 0 in the disordered phase (Gaussian fluctuations), 1 in the ordered phase (double-peak distribution), and takes an intermediate value in between. This value is non-universal and depends on the lattice size and shape, but at the critical temperature---where the magnetization scaling function is non-Gaussian \cite{teza2025universal}---the value is size-independent~\cite{BinderHeermannMC2010}.

We perform annealing runs for three system sizes---$N = 112$, $600$, and $2640$ spins---chosen to preserve an approximate aspect ratio $L_x/L_y \approx 1.5$ (Specifically, we used $(L_x,L_y)\in \{(14, 8), (30, 20), (60, 44)\}$ corresponding to $1.75$, $1.5$ and $\approx 1.36$). Each run samples spin configurations for a range of effective couplings $J$, as adjusted through our temperature-calibrated protocol. The resulting Binder cumulants are shown in Fig.~\ref{fig:binder_cumulant}.
Remarkably, the cumulant curves for all system sizes cross at a common coupling value $J_c =0.108\pm0.007$, strongly indicating that our sampling ensemble lies at the thermodynamic critical point.
The critical point at unit temperature is exactly known \cite{onsager1944crystal}, and an estimate of the critical temperature in the presence of a transverse field can be obtained through quantum Monte Carlo simulations \cite{PhysRevB.93.155157}.
Our sampling schedule provides direct control over the energy scales at play when measurements are performed, enabling us to provide direct estimates of physical and effective temperatures.
Therefore, we can provide here an estimate of the critical temperature for the 2D Ising model of $T_\textrm{crit}=2 k_B T_{\textrm{QPU}}^{\textrm{nom}}/(J_{\textrm{crit}}B(s^*))= 2.04 \pm 0.13$ under the presence of a transverse field $A(s^*)/(2k_B T_{\textrm{QPU}}^{\textrm{nom}})\approx 0.23$ ($k_B=1$ units), in line with quantum Monte Carlo predictions (see \cite{PhysRevB.93.155157}).
While deviations from the expected value could be ascribed to employment of superspins in the embedding (see Fig. \ref{fig:embedding}~\figpanel{b}), it is important to observe that the observed crossing value $b_N(J_\textrm{crit})=0.91 \pm 0.04$ closely matches Monte Carlo benchmarks for the 2D Ising model with periodic boundary conditions and $L_x/L_y = 1.5$~\cite{selke2006binder}, confirming that our annealing procedure produces correct critical fluctuations despite the hardware limitations that constrained the Binder cumulants' analysis in Ref. \cite{sathe2025classical}. 
To robustly estimate the uncertainties on $J_{\text{crit}}$ and $b_N(J_{\text{crit}})$, we perform a bootstrap resampling procedure consistent with the corresponding experimental uncertainties~\cite{Efron1994}.

This cumulant analysis provides a key validation of our sampling and embedding procedures, demonstrating that they yield thermal ensembles at effective temperature scales compatible with known classical criticality. The accuracy of the cumulant crossing further supports our choice of $s_f$ and the robustness of our in situ temperature correction strategy.

\section*{Critical Exponents and Finite-Size Scaling}
\noindent
To further characterize the universality class of the observed transition and quantify the consistency of our sampling protocol, we extract critical exponents from finite-size scaling of magnetization and susceptibility. We focus on the standard order parameter exponent $\beta$, susceptibility exponent $\gamma$, and correlation length exponent $\nu$ \cite{kadanoff2000statistical}.

We define the reduced coupling constant relative to the critical point as $\tau = (J_c - J)/J$ and express scaling hypotheses for observable quantities. The absolute magnetization density $\langle |m| \rangle = \langle |M| \rangle / N$ is expected to scale as 
\begin{equation}
\langle |m| \rangle = L^{-\beta/\nu} f_m(\tau L^{1/\nu}),
\end{equation}
and the susceptibility $\chi = N(\langle m^2 \rangle - \langle |m| \rangle^2)$ should obey
\begin{equation}
\chi = L^{\gamma/\nu} f_\chi(\tau L^{1/\nu}),
\end{equation}
where $f_m$ and $f_\chi$ are universal scaling functions, and $L = \sqrt{N}$ is the effective linear system size. These expressions follow from the finite-size scaling hypothesis, which posits that near criticality, observable quantities depend only on the combination $\tau L^{1/\nu}$~\cite{fisher1972scaling}. This framework underlies the expected collapse behavior for magnetization and susceptibility in finite systems~\cite{ballesteros1998finite}.

The critical exponents of the 2D Ising model are well-established, with exact values given by $\beta = \frac{1}{8}$, $\gamma = \frac{7}{4}$, and $\nu = 1$~\cite{onsager1944crystal}.
These values serve as a benchmark for validating the consistency of our sampling protocol and the fidelity of critical fluctuations.

Fig.~\ref{fig:exponents}~\figpanel{a} illustrates the scaling collapse for the three system sizes used in our Binder cumulant analysis. Panel~\figpanel{b} shows the magnetization collapse while panel~\figpanel{c} displays the susceptibility collapse~\cite{landau1976finite}. In both cases, we observe excellent collapse over a broad range of $\tau L^{1/\nu}$, especially near the critical region. The consistency and sharpness of the collapse across more than a decade in system size provide strong evidence that our sampling ensembles reflect equilibrium criticality.
To quantify this agreement and extract precise values of the critical exponents, we employed a technique to optimize the finite-size collapse \cite{Bhattacharjee2001AMO}.
Through a bootstrap analysis, we obtain the following exponents from the data: $\nu = 1.05 \pm 0.08$, $\beta = 0.09 \pm 0.01$, and $\gamma = 1.83 \pm 0.09$. These values show strong consistency with the theoretical values for the 2D Ising model~\cite{onsager1944crystal}, with $\nu$ and $\gamma$ in excellent agreement within $1\sigma$ of their theoretical values, and $\beta$
deviating only slightly from the expected value.
This agreement supports the identification of the observed transition with the Ising universality class. Importantly, the high quality of the data collapse represents a substantial improvement over earlier studies, which were often limited by smaller lattices, open boundary conditions, or uncontrolled thermal drift.

We emphasize that this scaling analysis is made possible by the consistency of the sampling protocol across both system sizes and annealing sessions together with the real-time temperature calibration routine which allows to remove systematic biases. The high quality of data collapse reflects not only the statistical fidelity of the samples but also the effectiveness of our protocol in stabilizing temperature and avoiding freeze-out effects.

\section*{Quantum Annealing Speedup Near Criticality}
\begin{figure}[t]
    \centering
    \includegraphics[width=\linewidth]{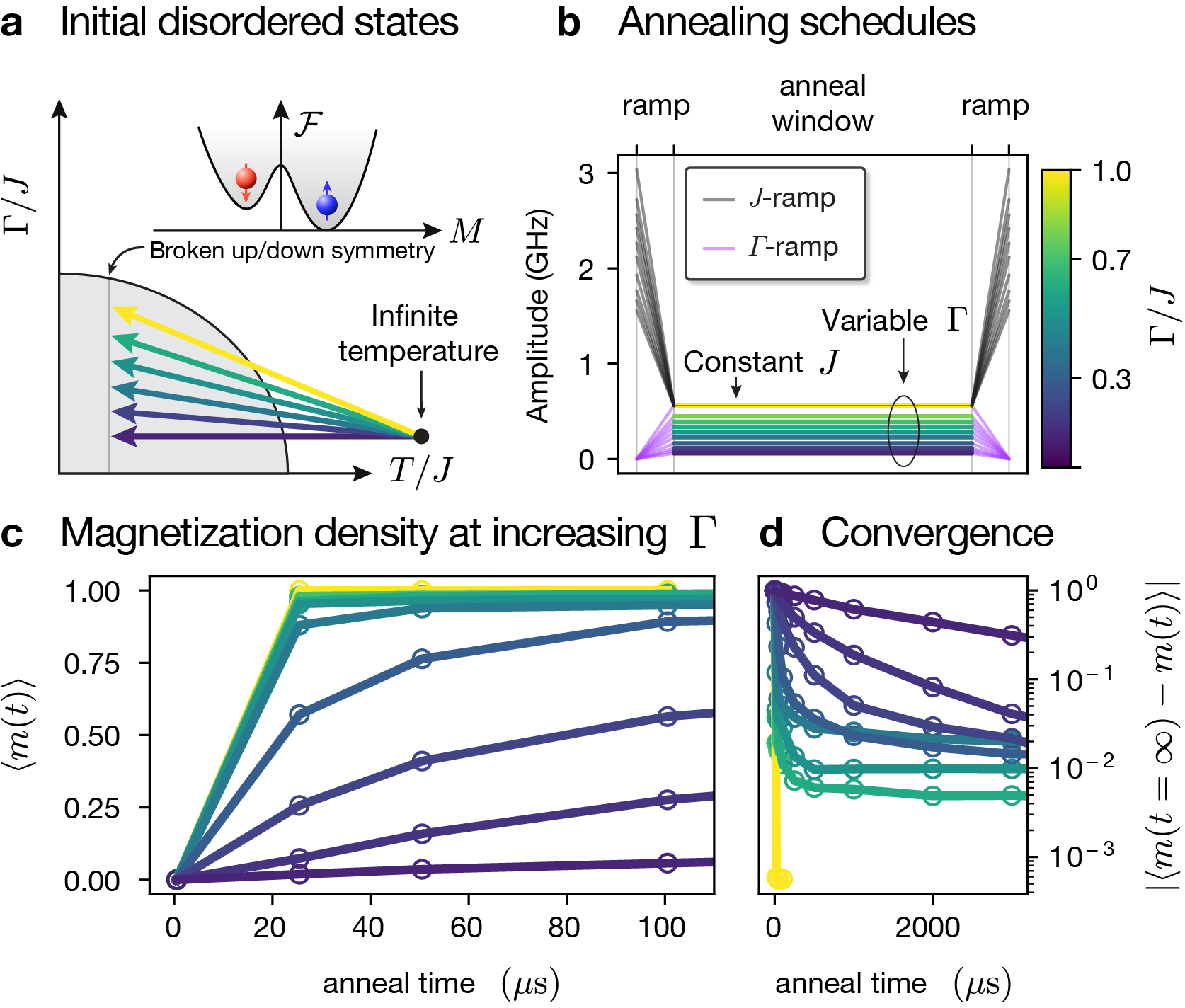}
    \caption{\textbf{Annealing speedup near criticality.}
    \textbf{\textsf{a}}, A schematic illustration of the experiment of panel b: a random configuration (corresponding to infinite temperature---paramagnetic phase) is quenched to Ising 2D Hamiltonian with various strength of the transverse field small enough to keep the system in the ferromagnetic phase.
    \textbf{\textsf{b}}, Anneal schedules to emulate evolutions at constant effective temperatures and different transverse fields amplitudes.
    Here $J/T\approx 0.9$, $\Gamma/J\in[0.1;1]$ and a small longitudinal field ($\epsilon=0.01$) breaks the symmetry favoring the positive direction.
    \textbf{\textsf{c}}, Magnetization density evolution $\left< m(t) \right>$ for several transverse field amplitudes $\Gamma$ (system size $N=2640$). Larger $\Gamma$ values lead to a faster convergence, fundamental to ensure steady-state convergence within the maximum annealing window.
    \textbf{\textsf{d}}, Convergence of the magnetization towards the steady state value $\left< m(t=\infty) \right>$. Failing to converge for small enough transverse field amplitudes is related to single realizations thermalizing in the wrong (metastable) well.
    Therefore, larger $\Gamma$ values increase reliability in the convergence to the symmetry-broken ground state. 
    }
    \label{fig:advantage_mag_conv}
\end{figure}

\noindent
Quantum annealing (QA) is often benchmarked by its ability to reach true ground states in complex energy landscapes~\cite{MandraExponentially2017,PhysRevE.105.035305}, where metastable states separated by large barriers challenge quantum and classical methods alike.
While a full understanding of the transition towards a spin-glass phase is still missing \cite{singh2017critical,miyazaki2013realspace}, a recent work highlighted how, in the presence of certain symmetries, quantum annealing can still provide an advantage over classical methods when crossing a critical line \cite{bernaschi2024quantum}, both in terms of accuracy and anneal time.

Here, we test QA's ability to resolve such barriers and escape from non-equilibrium conditions by initializing the system in either disordered or metastable configurations.
In both cases, we consider a transverse-field Ising Hamiltonian with a longitudinal bias field,
\begin{equation}\label{Eq:TransFieldHamilt}
\hat{H} = -J \left( \sum_{\langle i,j \rangle} \sigma^z_i \sigma^z_j +\epsilon \sum_i \sigma^z_i\right)- \Gamma \sum_i \sigma^x_i,
\end{equation}
where $J$ is the Ising interaction strength between nearest-neighbor spin pairs $\langle i,j\rangle$, $\Gamma$ is the transverse field, and $J\epsilon$ is a uniform longitudinal field that biases the system toward a symmetry-broken state.

First, we investigate how a disordered spin configuration evolves under a fixed Hamiltonian with nonzero transverse and longitudinal fields. This setup mimics a quench from an infinite-temperature state to a low-temperature effective Hamiltonian, as shown in Fig.~\ref{fig:advantage_mag_conv}~\figpanel{a}, allowing us to probe the system’s relaxation dynamics and the nature of its final state. To do so, we initialize the system in a random spin configuration and quench to a transverse-field Ising Hamiltonian with fixed $\Gamma$ and $\epsilon = 0.01$. The system was allowed to evolve for varying durations, and the average longitudinal magnetization was recorded as a function of time. This experiment was repeated for several values of the transverse field $\Gamma$, as shown in Fig.~\ref{fig:advantage_mag_conv}~\figpanel{b}.

The results, shown in Fig.~\ref{fig:advantage_mag_conv}~\figpanel{c},~\figpanel{d}, indicate that at small transverse fields, the relaxation is slow and the magnetization saturates at a value significantly below the equilibrium expectation. This indicates that the system often becomes trapped in metastable states, failing to reach the true symmetry-broken minimum on the timescale of the experiment. As the transverse field increases, relaxation becomes faster and the saturation magnetization approaches the expected ground-state value.

\begin{figure}[t]
    \centering
    \includegraphics[width=\linewidth]{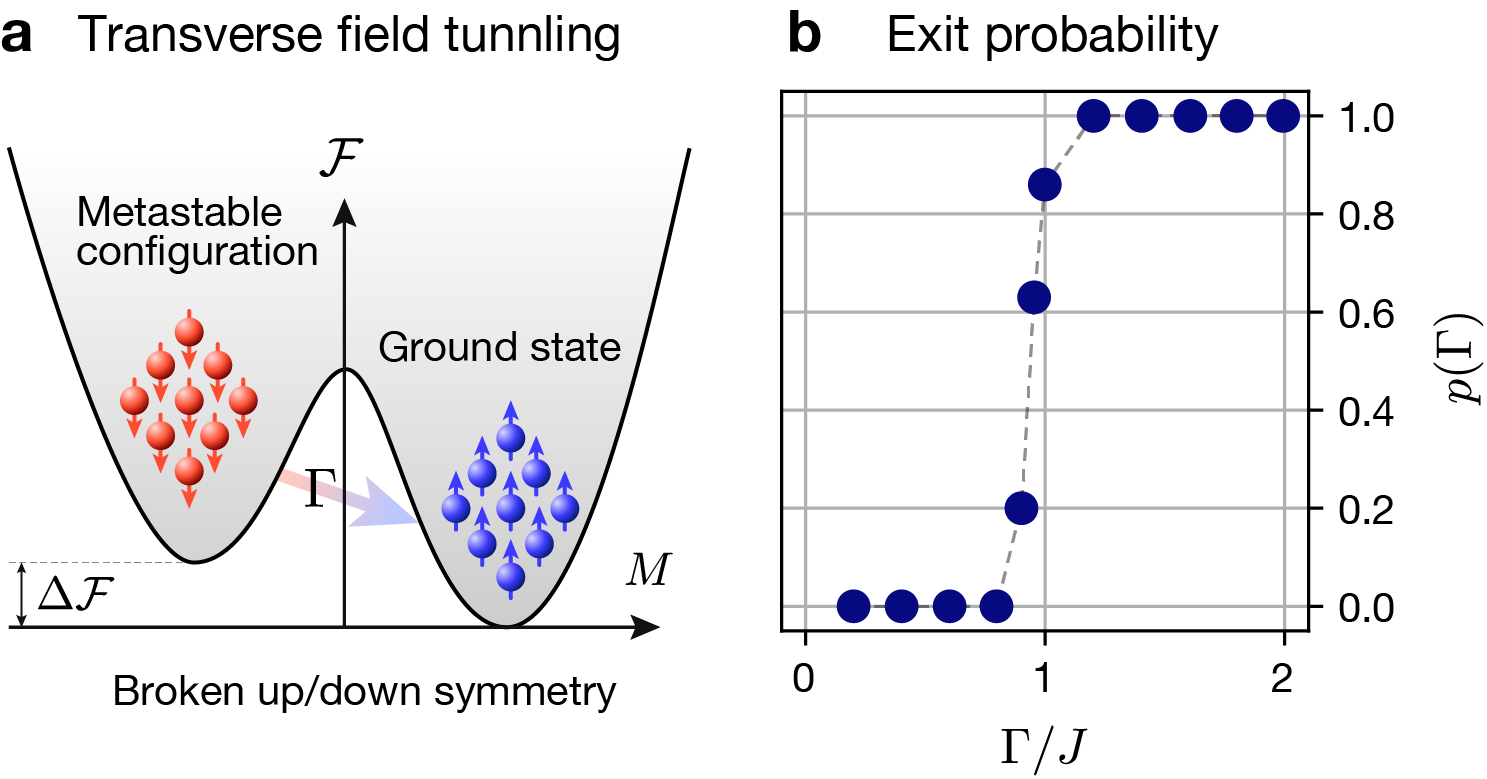}
    \caption{\textbf{Barrier crossing via transverse field.}
    \textbf{\textsf{a}}, A schematic illustration of the experiment of panel b: the system is prepared in the metastable configuration (corresponding to all-down spins) of the Ising 2D Hamiltonian (Eq. \ref{Eq:TransFieldHamilt}), with $T/J\approx 1.1$, $\epsilon=0.05$ (system size $N=2640$).
    Increasing the transverse field amplitude---while staying in the ferromagnetic phase ( $\Gamma/J\in[0.2;2]$)---should eventually enable the crossing of the barrier and lead the system to the correct groundstate.
    \textbf{\textsf{b}}, Probability $p(\Gamma)$ of successful reversal from the metastable (all-down) configuration to the symmetry-breaking ground state as a function of $\Gamma$. The sharp change indicates an enhanced ability to escape metastable states due to {the transverse field}.}
    \label{fig:advantage_exit}
\end{figure}

We then consider a setting in which the system is initialized in a fully polarized state opposite to the direction favored by the longitudinal field, i.e., a metastable configuration, as illustrated in Fig.~\ref{fig:advantage_exit}~\figpanel{a}. The system is quenched to the transverse-field Ising Hamiltonian of Eq.~\ref{Eq:TransFieldHamilt}, with a fixed longitudinal bias of $\epsilon = 0.05$ and varying values of the transverse field $\Gamma$. After a hold time of $1000~\mu$s, the spin configuration is measured, and the probability of reaching the correct symmetry-broken ground state is recorded as a function of $\Gamma$.

The results, shown in \ref{fig:advantage_exit}~\figpanel{b}, indicate that at for $\Gamma/J\ll1$ the system remains trapped in the metastable state within the anneal timescale.
{In this regime, the microscopic quantum dynamics of magnetization domains is perturbatively coherent~\cite{balducci2022localization,balducci2023interface}, supporting the stability we observe.}
As $\Gamma$ increases, the success probability rises sharply around $\Gamma/J=1$ , signaling the onset of barrier crossing. This transition reflects the role of {the transverse field} in facilitating escape from metastability.
Plugging the experimental parameters ($J = 0.3212$~GHz, $\epsilon = 0.05J$) with the observation that the transition happens at $\Gamma\approx 1$, implies that the system is able to tunnel through a barrier height of approximately $80~\mu$eV, which is roughly 60 times larger than the thermal energy of 17~mK.

Let us note that the values of $\Gamma$ explored in the experiment remain well below the critical transverse field $\Gamma_c$ at which the ordered phase disappears and the free energy barrier vanishes altogether \cite{friedman1978ising}. The zero-temperature quantum critical point for the transverse-field Ising model on the square lattice occurs at $\Gamma_c/J \approx 3.044$, and in our experiment, where  $J/T_\textrm{QPU}\approx 1.1$ the critical transverse field is bounded by $\Gamma_c/J> 2.5$  \cite{PhysRevB.93.155157}. Here, the maximum value of $\Gamma$ used corresponds to $\Gamma/J \approx 2$, and the effective temperature remains well below the thermal phase boundary. Therefore, we are operating in a regime where symmetry-broken states remain metastable minima of the free energy landscape, and the concept of a barrier between them is still well-defined. This ensures that the observed transition from trapping to escape reflects a meaningful crossing (or tunneling) event between locally stable configurations, rather than a smooth crossover between disordered states.

\section*{Conclusions}
\noindent
In this work, we demonstrated that quantum annealers, when carefully calibrated, can accurately resolve finite-temperature criticality in complex many-body systems carefully. The calibration of QPU temperature and mitigation of freeze-out effects proved essential to reliably extracting critical temperature and exponents of the 2D Ising ferromagnet.
While hardware limitations (native topology of the QPU, freeze-out point) have been regarded so far as bottlenecks---even in more recent architectures~\cite{dwave_docs,sathe2025classical}---our operational line bypasses these limitations, enabling reliable analysis for finite-temperature criticality both in current and older architectures.
Thanks to our embedding strategy, our approach extends to a broad class of systems and problems that map to the Ising model. This includes quantum gravity models~\cite{Livine2024,Sahay2025}, non-equilibrium quantum thermodynamics, e.g., charging and work extraction in quantum Ising networks~\cite{Campaioli2024, Donelli2025}, anomalous relaxation phenomena~\cite{Teza2025}, e.g., quantum Mpemba effect~\cite{Kochsiek2022}, and quantum circuit compilation~\cite{Rattacaso2024}. Further work on this approach should address the issue of memory effects (non-Markovianity) that occur between quantum annealing schedules~\cite{Zambon2025}. Finally, systematic comparisons with classical algorithms could further clarify the nature and extent of any potential advantage of quantum sampling or optimization near criticality.
\vspace{5pt}

\section*{Acknowledgments}
\noindent
G. T. acknowledges support by the Max Planck Society. O. R. acknowledges financial support from the ISF, grant no. 232/23, and of the Minerva Stiftung. All the authors acknowledge the quantum computing resources provided by the Quantum Computing and Simulation Center (QCSC) and CINECA.
We thank Daniele Ottaviani for technical support on the quantum computing resources, we also thank Francesco Caravelli for insightful discussions and remarks.

\bibliographystyle{apsrev4-2}
\bibliography{references}

\end{document}